\newcommand{\PSD}{\hat{\psi}^\dagger}
\newcommand{\PSI}{\hat{\psi}}
\newcommand{\rv}{({\bf r})}

\documentstyle[prl,aps,twocolumn,epsf]{revtex}

\begin{document}

\draft

\title
{
Vortex stabilization in Bose-Einstein condensate of alkali atom gas
}
\author
{
Tomoya Isoshima\cite{tomoya} and Kazushige Machida
}
\address{
Department of Physics, Okayama University, Okayama 700-8530, Japan
}

\date{Submitted}

\maketitle

\begin{abstract}
A quantized vortex in the Bose-Einstein condensation (BEC),
which is known to be unstable intrinsically,
is demonstrated theoretically to be stabilized
by the finite temperature effect.
The mean-field calculation of Popov approximation
within the Bogoliubov theory is employed,
giving rise to a self-consistent solution for 
BEC confined by a harmonic potential.
Physical origin of this vortex stabilization is investigated.
An equivalent effect is also proved to be induced
by an additional pinning potential at the vortex center
produced by a focused laser beam even at the lowest temperature.
The self-consistent solutions give detailed properties of a stable vortex,
such as the spatial profiles of the condensate and non-condensate,
the particle current density around the core,
the whole excitation spectrum, and their temperature dependences.
\end{abstract}


\section{Introduction}
Many-body dilute Bose systems are the focus in condensed matter physics
since the experimental realization of Bose-Einstein condensation (BEC) 
in alkali atom gases at ultra-low temperatures\cite{cornell,hulet,ketterle}.
There is a pile of recent
theoretical and experimental works
devoted to elucidation of BEC~\cite{review2,review1}.
Experimental works are mainly conducted for dilute Bose systems,
whose atom number is typically $\sim O(10^6)$ for $^{23}$Na and $^{87}$Rb.
Gases are confined magnetically in a harmonic trap.
The BEC transition temperatures are in a range of $O(\mu \text{K})$.
The experimental forefront is rapidly extending;
one can tune the particle-particle interaction constant by
utilizing the so-called Feshbach resonance phenomena\cite{feshbach},
or one can confine a system purely optically so that
several atomic hyperfine substates are simultaneously condensed,
giving rise to an opportunity to explore a BEC
with internal degrees of freedom~\cite{kurn,ohmi,ho}.
Several proposals of experimental method
to realize a vortex state have made
theoretically\cite{review1,laserkarl,gp2d}.
This paper concerns the vortex states in somewhat different way.

As for the earlier theoretical developments,
the microscopic theoretical work on BEC has started with
Bogoliubov\cite{bogoliubov} long time ago
and followed by several important progresses, 
such as Gross\cite{gross}, Pitaevskii\cite{pitaevskii},
Iordanskii\cite{iordanskii}, Fetter\cite{fetter}, and Popov\cite{popov}.
These mean-field theories are particularly suited for treating
the present gaseous BEC systems
trapped optically or magnetically in a restricted geometry. 
Therefore, the current theoretical studies\cite{meanfield} mainly
focus on examining these mean field theories
to extract the spatial structures of the condensate and non-condensate
and low-lying collective modes.
Agreement between these mean field theories and experiments
is fairly good\cite{review1}.

Previously we have demonstrated that
Bogoliubov approximation (BA) of the mean-field theory
cannot sustain a stable vortex\cite{isoshima}.
This somewhat unexpected result,
which had been hinted by Dodd\cite{dodd},
is now clearly explained physically by Rokhsar\cite{rokhsar}
who shows that the vortex is always unstable against the
translational mode with zero energy.
Thus the vortex core in a cylindrically-symmetric 
harmonic trap spirals out towards the outside of a system where
the condensate density is low.

The main purpose of this paper is
to examine the stability of a quantized vortex
which is directly related to the question of superfluidity
of the present BEC systems.
This instability can be expressed numerically
either by comparing the free energy of systems
with and without a vortex (stability), or
by checking whether an eigenstate with a negative eigenvalue
exists or not in a system with vortices (metastability).
The latter is far serious than the former and
is discussed in this paper as ``stability".

In the present paper,
which follows our series of the papers\cite{isoshima,isoshima2} 
based on the same formalism and numerical procedure,
we show that the intrinsic instability of a quantized vortex
can be remedied either by raising temperature ($T$) or
by introducing the pinning potential at the center
of a cylindrically symmetric rectilinear vortex line,
which prevents the vortex from spiraling out.
The latter method for the vortex stabilization is pointed out
by several authors\cite{review1},
whose experimental feasibility seems to be quite high
since it may be easy to realize a desired pinning potential
by sending a focused laser beam at the vortex center.
Here we provide microscopic self-consistent calculations
within the Popov approximation, 
yielding detailed properties of a stable vortex,
beyond Gross-Pitaevskii theory
which treats only the condensate component.
This enables us to discuss at finite temperatures.

Various spatial structures of
the quantized vortex in rotating BEC systems
are investigated based on this scheme of stabilization. 
This investigation is important since the microscopic understanding
of the vortex
based on the self-consistent picture
has been lacking so far,
except for a few attempts\cite{gpvortex,sinha,svi} for present BEC systems.
Several points are worth exploiting:
(1) the spatial profiles of the condensate and non-condensate,
(2) the dispersion relation of the excitation spectrum of the system,
and (3) the low-lying excitations localized around a vortex core,
which is known as the Kelvin mode\cite{donnelly}.
These features may be directly observable and should
be compared with those in vortices in a superconductor
where extensive studies have been compiled\cite{hohen}.

Arrangement of this paper is as follows:
In the next section we recapitulate the Popov approximation (PA)
which is one of various types of mean-field theories (see for example,
Refs.~\cite{isoshima,hutchinson})
due originally to Bogoliubov\cite{bogoliubov}.
We explain our numerical procedure
to solve a self-consistent equation within the PA.
In Sec. 3
the physical origin of stabilization is discussed and
the detailed calculations for a stable vortex
at finite temperatures are shown.
We also compare the calculated vortex systems
to the vortex state in type-II superconductors.
The pinning potential is introduced to yield a stable vortex in Sec. 4
where the strength and width of the potential are critically
examined for the vortex stabilization.
We devote the final section to discussions and conclusion.

\section{Formulation and Numerical Procedure}
\subsection{Mean Field Approximation}

We start with the following Hamiltonian in which 
Bose particles interact with a two-body potential:
\begin{eqnarray}
  \hat{\text{H}} &=& \int\!\!d{\bf r}
  \hat{\Psi}^{\dagger}\rv 
  \left(
    - \frac{\hbar^2 \nabla^2}{2m} + V\rv - \mu
  \right) \hat{\Psi}\rv             \nonumber
\\ & &
  + \frac{g}{2} 
  \int\!\!d{\bf r}
  \hat{\Psi}^{\dagger}({\bf r}) \hat{\Psi}^{\dagger}({\bf r}) 
  \hat{\Psi}({\bf r}) \hat{\Psi}({\bf r})  \label{eq:h1}
\end{eqnarray}
where the chemical potential $\mu$ is introduced to fix the particle number
and $V\rv$ is the confining potential.
The two-body interaction between an atom at ${\bf r}_1$ and
one at ${\bf r}_2$ is assumed to
be $g\delta({\bf r}_1 - {\bf r}_2)$ with
$g$ being a positive (repulsive) constant
proportional to the s-wave scattering length $a$,
namely $g=4\pi \hbar^2a/ m$ ($m$ is the particle mass).

In order to describe the Bose condensation,
we assume that the field operator $\hat{\Psi}$ is decomposed into
\begin{equation}
  \hat{\Psi}\rv = \phi\rv + \PSI\rv \label{eq:deco}
\end{equation}
where the ground state average is given by
\begin{equation}
  \langle \hat{\Psi}\rv \rangle = \phi\rv .
\end{equation}
A c-number $\phi\rv$ corresponds to the condensate wave function
and $\PSI\rv$ is a q-number describing the non-condensate.
Substituting the above decomposition (\ref{eq:deco}) in (\ref{eq:h1}),
we obtain
\begin{eqnarray}
  \hat{\text{H}} &=&
    \int d{\bf r}\Bigl[
      \phi^*\rv \left\{
        h\rv \phi\rv +{1\over 2}g|\phi\rv |^2
      \right\}\phi\rv     \nonumber
\\ & &
      + \PSD\rv \left\{
        h\rv + g |\phi\rv |^2
      \right\} \phi\rv
      + \text{h.c.}          \nonumber
\\ & &
      + \PSD\rv
        \left\{ h\rv + 2g|\phi\rv |^2 \right\}
      \PSI\rv          \nonumber
\\ & &
      + \frac{g}{2}\PSD\rv \PSD\rv \phi\rv \phi\rv 
      + \frac{g}{2}\PSI\rv \PSI\rv \phi^*\rv \phi^*\rv      \nonumber
\\ & &
      + g\PSD\rv \PSI\rv \PSI\rv \phi^*\rv
      + g\PSD\rv \PSD\rv \PSI\rv \phi\rv  \nonumber
\\ & &
      + \frac{g}{2}\PSD\rv \PSD\rv \PSI\rv \PSI\rv
    \Bigr]           \label{eq:h}
\end{eqnarray}
where $h\rv \equiv -\frac{\hbar^2\nabla^2}{2m} + V\rv - \mu $ is
the one-body Hamiltonian.
Let us introduce the variational parameter: 
the non-condensate density $\rho\rv = \langle \PSD \PSI \rangle$ 
and approximate as 
$\PSD \PSI \PSI \sim 2\PSI \rho$, and
$\PSD \PSD \PSI \PSI \sim 4\PSD \PSI \rho.$

Then, $\hat{\text{H}}$ is rewritten as
\begin{eqnarray}
  \hat{\text{H}} &=&
    \int d{\bf r}
    \Bigl[
      \phi^*\rv \left\{
        h\rv \phi\rv
      + {1\over 2}g|\phi\rv |^2
      \right\}\phi\rv            \nonumber
\\ & &
      + \PSD\rv \left\{
        h\rv + g|\phi\rv |^2 + 2g\rho\rv
      \right\}\phi\rv
      + \text{h.c.}       \nonumber
\\ & &
      + \PSD\rv
      \left\{
        h\rv + 2g(|\phi\rv |^2 + \rho\rv ) 
      \right\}
      \PSI\rv          \nonumber 
\\ & &
      + \frac{g}{2}\PSD\rv \PSD\rv \phi\rv   \phi\rv 
      + \text{h.c.}
    \Bigr].
\end{eqnarray}
In order to diagonalize this Hamiltonian, the following Bogoliubov
transformation is employed, namely, 
$\PSI\rv$ is written in terms of 
the creation and annihilation operators $\eta _q$ and $\eta _q^{\dagger}$ 
and the non-condensate wave functions $u_q\rv$ and $v_q\rv$ as
\begin{equation}
  \PSI\rv = \sum _{{\it q}}
    \left[
      u_q\rv \eta_q - v_q^*\rv \eta_q^{\dagger}
    \right]   \label{eq:psieta}
\end{equation}
where $q$ denotes a set of the quantum numbers.
This leads to the diagonalized form:
 $\hat{\text{H}} =
\text{E}_0 + \sum_q \varepsilon_q \eta _q^{\dagger}\eta _q.$
The condition that the first order term in
$\PSI\rv$ vanish yields
\begin{equation}
  \left[
    h\rv + g |\phi\rv |^2 + 2g\rho\rv
  \right] \phi\rv = 0.  \label{eq:gp}
\end{equation}
When $\rho\rv $ is made zero, 
it reduces to the commonly used Gross-Pitaevskii equation
\begin{equation}
  \left[
    h\rv + g|\phi\rv |^2 
  \right] \phi\rv = 0,  \label{eq:gpo}
\end{equation}
which is a non-linear Schr\"odinger type equation.

The condition that the Hamiltonian be diagonalized gives rise to 
the following set of eigenvalue equations for $u_q\rv$ and $v_q\rv$ with 
the eigenvalue $\varepsilon_q$:
\begin{eqnarray}
  \left[
    h\rv + 2g \{ |\phi\rv |^2 + \rho\rv \}
  \right] u_q\rv   \nonumber
& & \\
  - g\phi^2\rv v_q\rv    
  &=&
  \varepsilon_q u_q\rv          \label{eq:bg1}
\\
  \left[
    h\rv + 2g \{ |\phi\rv |^2 + \rho\rv \}
  \right] v_q\rv \nonumber
& & \\
  - g \phi^{*2}\rv u_q\rv
  &=&
  -\varepsilon_qv_q\rv .   \label{eq:bg2}
\end{eqnarray}
The chemical potential $\mu$ is contained in $h$ 
of the left hand side of Eqs.\ (\ref{eq:bg1}) and (\ref{eq:bg2}).
Therefore $\mu$ corresponds to the origin of the energy for
each eigenvalue $\varepsilon_q$.

The eigenfunctions
$u_q\rv$ and $v_q\rv$ must satisfy the normalization condition:
\begin{equation}
  \int \left[
    u_p^*\rv u_q\rv - v_p^*\rv v_q\rv
  \right] d{\bf r}
  = \delta _{p,q}. \label{eq:nor}
\end{equation}
This also means that 
$\int |u_q|^2 d{\bf r}$ is larger than $\int |v_q|^2 d{\bf r}$
for an arbitrary $q$.
The variational parameter
$\rho\rv$ is determined self-consistently by 
\begin{eqnarray}
  \rho\rv &=&
    \langle \PSD \PSI \rangle \nonumber
\\
  &=&
    \sum_{\it q} \left[
      |u_q\rv |^2 f(\varepsilon_q)
      + |v_q\rv |^2 (f(\varepsilon_q) + 1)
    \right], \label{eq:rho}
\end{eqnarray}
where $f(\varepsilon)$ is the Bose distribution function.
Equations (\ref{eq:gp}), (\ref{eq:bg1}), (\ref{eq:bg2}), (\ref{eq:nor}), and 
(\ref{eq:rho}) constitute a complete set of the 
self-consistent equations for the Popov approximation (PA).
The iterative calculations of these equations yield a convergent
self-consistent solution.

If the non-condensate component $\rho(r)$ is neglected in this set of
equations,
the resulting set of the equations yields the
so-called Bogoliubov approximation (BA).
This was extensively discussed analytically by Pitaevskii\cite{pitaevskii},
Iordanskii\cite{iordanskii}, and Fetter\cite{fetter}.

The expectation value of the particle number density is given as
\begin{eqnarray}
  \langle \hat{\text{n}}\rv \rangle = |\phi\rv |^2 + \rho\rv ,
\end{eqnarray}
that is, the total density consists of the condensate part $\phi\rv$
and the non-condensate part $\rho\rv$.
The particle current density is calculated as
\begin{eqnarray}
  {\bf j}\rv &=&
    \frac{\hbar}{2mi}\{
      \phi^*\rv \nabla\phi\rv - \phi\rv \nabla\phi^*\rv
    \}  \nonumber
\\
  & &+
    \frac{\hbar}{2mi}\langle 
      \PSD\rv \cdot \nabla \PSI\rv - \nabla \PSD\rv \cdot \PSI\rv
    \rangle.            \label{eq:j}
\end{eqnarray}
The local density of states is calculated as
\begin{eqnarray}
  N(E, r) = 
    \sum_{\bf q} \left\{
      |u_{\bf q}(r)|^2 + |v_{\bf q}(r)|^2 
    \right\}
    \delta(\varepsilon_{\bf q}-E).       \label{eq:dos}
\end{eqnarray}
This quantity was observed directly by scanning tunneling microscopy
for a vortex in superconductors\cite{hess}.

\subsection{Vortex Description}

We now consider a cylindrically symmetric system
which is characterized by the radius $R$ and the height $L$.
We use the cylindrical coordinate: ${\bf r} = (r,\theta ,z)$.
We impose the boundary conditions that
all the wave functions vanish at the wall $r=R$, which
is taken far enough from the vortex center at $r=0$, and
the periodic boundary condition along the $z$-axis.
When a vortex line passes through the center of the cylinder,
the condensate wave function $\phi\rv$ is expressed as 
\begin{equation}
  \phi(r,\theta ,z) = \phi(r) e^{iw\theta } 
\end{equation}
where $\phi(r) $ is a real function and $w$ is the winding number.
$w=0$ corresponds to the non-vortex case and $w=1$ to the vortex case.
The $w\ge 2$ case is not considered here
because this state is energetically too unstable.
The non-condensate density $\rho\rv$ is a real function,
depending only on $r$, that is, $ \rho (r,\theta ,z) = \rho(r)$.
It is also seen from Eqs.\ (\ref{eq:bg1}) and (\ref{eq:bg2}) that 
the phases of $u_q\rv$ and $v_q\rv$ are written as 
\begin{eqnarray}
  u_{\bf q}\rv &=&
    u_{\bf q}(r)e^{iq_zz}e^{i(q_{\theta } + w)\theta} \label{eq:phase:u}
\\
  v_{\bf q}\rv &=&
    v_{\bf q}(r)e^{iq_zz}e^{i(q_{\theta } - w)\theta}. \label{eq:phase:v}
\end{eqnarray}
The set of the quantum numbers ${\bf q}$ in (\ref{eq:psieta})
is described by $(q_r, q_{\theta } , q_z)$ where
$q_r = 1, 2, 3, \cdots $, $q_\theta = 0, \pm 1, \pm 2, \cdots$, 
$q_z = 0, \pm 2\pi /L, \pm 4\pi /L, \cdots$.
Note that only $u(r)(q_{\theta}=-1)$ and
$v(r)(q_{\theta}=1)$ are non-vanishing at $r=0$.

By following the method by Gygi and Schl\"uter\cite{gygi},
the functions $u_q(r)$ and $v_q(r)$ are expanded
in terms of 
\begin{equation}
  \varphi_{\nu}^{(i)}(r) \equiv
    \frac{\sqrt{2}}{
      J_{|\nu| +1}
      \left(
        \alpha_{|\nu|}^{(i)}
      \right)
    }
  J_{|\nu|}\left(\alpha_{|\nu|}^{(i)}\frac{r}{R}\right)
\end{equation}
as
\begin{eqnarray}
  u_q(r) &=&
    \sum _i c_q^{(i)}\varphi _{q_{\theta }+w}^{(i)}(r) \\
  v_q(r) &=&
    \sum _i d_q^{(i)}\varphi _{q_{\theta }-w}^{(i)}(r)
\end{eqnarray}
where $J_{\nu}(r)$ is the Bessel function of $\nu $-th order and
$\alpha_{\nu}^{(i)}$ denotes $i$-th zero of $J_{\nu}$.

The eigenvalue problem of Eqs.\ (\ref{eq:bg1}) and (\ref{eq:bg2})
are reduced to diagonalizing the matrix:
\begin{equation}
  \left( \!\!
    \begin{array}{cc}
      A_{i,j}(q_{\theta}+w,q_z) & -B_{i,j}(q_{\theta},w)   \\
      & \\
      B_{i,j}^{\text{T}}(q_{\theta},w) & -A_{i,j}(q_{\theta}-w,q_z)
    \end{array} \!\!
  \right) \!\!\!
  \left( \!\!
     \begin{array}{c}
       c_{\bf q}^{(1)}  \\ c_{\bf q}^{(2)}  \\ \vdots    \\
       d_{\bf q}^{(1)}  \\ d_{\bf q}^{(2)}  \\ \vdots
     \end{array} \!\!
  \right)\!\!
  = \varepsilon_{\bf q} \!\!
  \left( \!\!
    \begin{array}{c}
      c_{\bf q}^{(1)}   \\ c_{\bf q}^{(2)}   \\ \vdots    \\
      d_{\bf q}^{(1)}   \\ d_{\bf q}^{(2)}   \\ \vdots 
    \end{array} \!\!
  \right) \label{eq:c:mat}
\end{equation} 
for each $q_{\theta}$ and $q_z$, where 
\begin{eqnarray}
  A_{i,j}(\nu ,q_z) 
  &=&
  \frac{\hbar^2}{2m}
  \left\{
    \left(\frac{\alpha _{\nu }^{(j)}}{R} \right)^2 + q_z^2
  \right\}\delta_{i,j} - \mu\delta_{i,j}      \nonumber
\\ & &
  + \int_0^R
    \left( V + 2g (\phi^2 + \rho) \right)
    \varphi_{\nu }^{(i)}\varphi _{\nu }^{(j)}
  r\, dr       \label{eq:c:mat:a}
\\
  B_{i,j}(q_{\theta },w) 
  &=&
  \int _0^R
    g\phi ^2 
    \varphi _{q_{\theta }+w}^{(i)}\varphi _{q_{\theta }-w}^{(j)} 
  r\, dr. \label{eq:c:mat:b}
\end{eqnarray}

The following symmetry relation should be noted:
\begin{equation}
  (c, d, \varepsilon) \text{ for } q_{\theta}
\Leftrightarrow
  (d, c, -\varepsilon) \text{ for } -q_{\theta}.
\end{equation}
We determined the signs of $q_{\theta}$ and $\varepsilon$ by the
normalization condition Eq.\ (\ref{eq:nor}).

It is not self-evident a priori that 
Eq.~(\ref{eq:c:mat}) gives real eigenvalues because
the Hamiltonian matrix in Eq.\ (\ref{eq:c:mat})
is not symmetric.
However, these eigenvalues can be proved to be real (positive or negative)
by a way similar to that of Fetter\cite{fetter}.

The circulating current density $j_{\theta}(r)$
in Eq.\ (\ref{eq:j}) is expressed as 
\begin{eqnarray}
  j_{\theta}(r)&=&j^{(1)}_{\theta}(r)+j^{(2)}_{\theta}(r) \label{eq:j0}
\\
  j^{(1)}_{\theta}(r)&=&{\hbar\over m}{w \over r}\cdot \phi^2(r) \label{eq:j1}
\\
  j^{(2)}_{\theta}(r) &=&
    {\hbar\over m}\sum_{\bf q}
    \Bigl\{
      {q_{\theta}+w \over r}|u_{\bf q}(r)|^2f(\varepsilon_{\bf q})  \nonumber
\\ & &
     -{q_{\theta}-w \over r}|v_{\bf q}(r)|^2(f(\varepsilon_{\bf q})+1)
    \Bigr\} \label{eq:j2}
\end{eqnarray}
where the total current consists of the condensate component
$j^{(1)}_{\theta}(r)$ and the non-condensate $j^{(2)}_{\theta}(r)$.

\subsection{Calculated System}

We have performed self-consistent calculations
of a gas of Na atoms trapped radially
by a harmonic potential $V(r) = \frac{1}{2}m(2\pi \nu_r)^2 r^2$.
We use the following parameters:
$m = 3.81 \times 10^{-26} \text{kg}$ ,
the scattering strength $a = 2.75 \text{nm}$,
and the radial trapping frequency $\nu_r = 200 \text{Hz}$.
The area density per unit length along the z-axis is chosen to be
$n_z = 2\times 10^4 /\mu \text{m}$ ,
so that the peak density will be $O(10^{20} \text{m}^{-3})$. 
The system size is set to $R=20 \mu \text{m}$ and $L = 10 \mu \text{m}$.

We have employed the energy cutoff method
to calculate
Eqs.~(\ref{eq:rho}) and (\ref{eq:j2})
where the calculation terminates when the obtained
eigenvalues exceeds about $80 h\nu_r$.
This method treats $\sim 80000$ eigenfunctions,
although it is lowest two modes,
namely the condensate state and the lowest core localized state (LCLS),
that play significant role.

\section{Stable Quantized Vortex at Finite Temperatures}

The isolated single quantized vortex is unstable within certain mean-field
approximations\cite{isoshima,rokhsar}
in a non-rotating system.
When the lowest excitation eigenvalue of the system
becomes lower than the energy of the condensate,
the whole calculation based on the assumption that
the energy for the condensate is lowest\cite{isoshima} is invalidated. 
This means that the vortex is unstable.

In the finite $T$ calculations shown below
the quantized vortex becomes stable in the BEC systems
confined by a harmonic potential above a certain temperature.
We analyze how the vortex state at finite $T$
can become stabilized,
and examine fundamental properties of a stable vortex.

\subsection{Spatial Structures of Condensate and Non-condensate}

\begin{figure}
\epsfxsize=7cm
\epsfbox{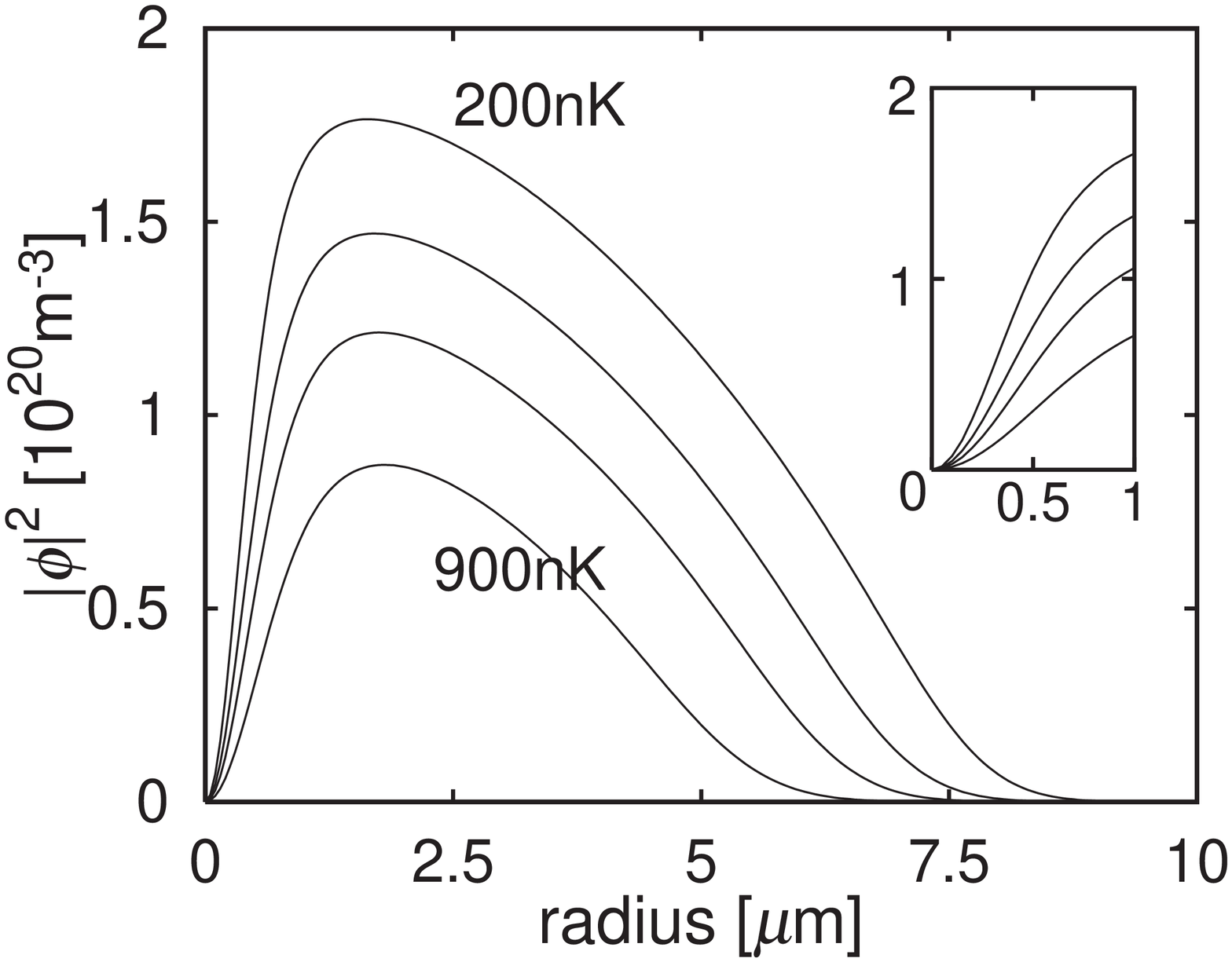}
\hspace{3.5cm}(a)

\epsfxsize=7cm
\epsfbox{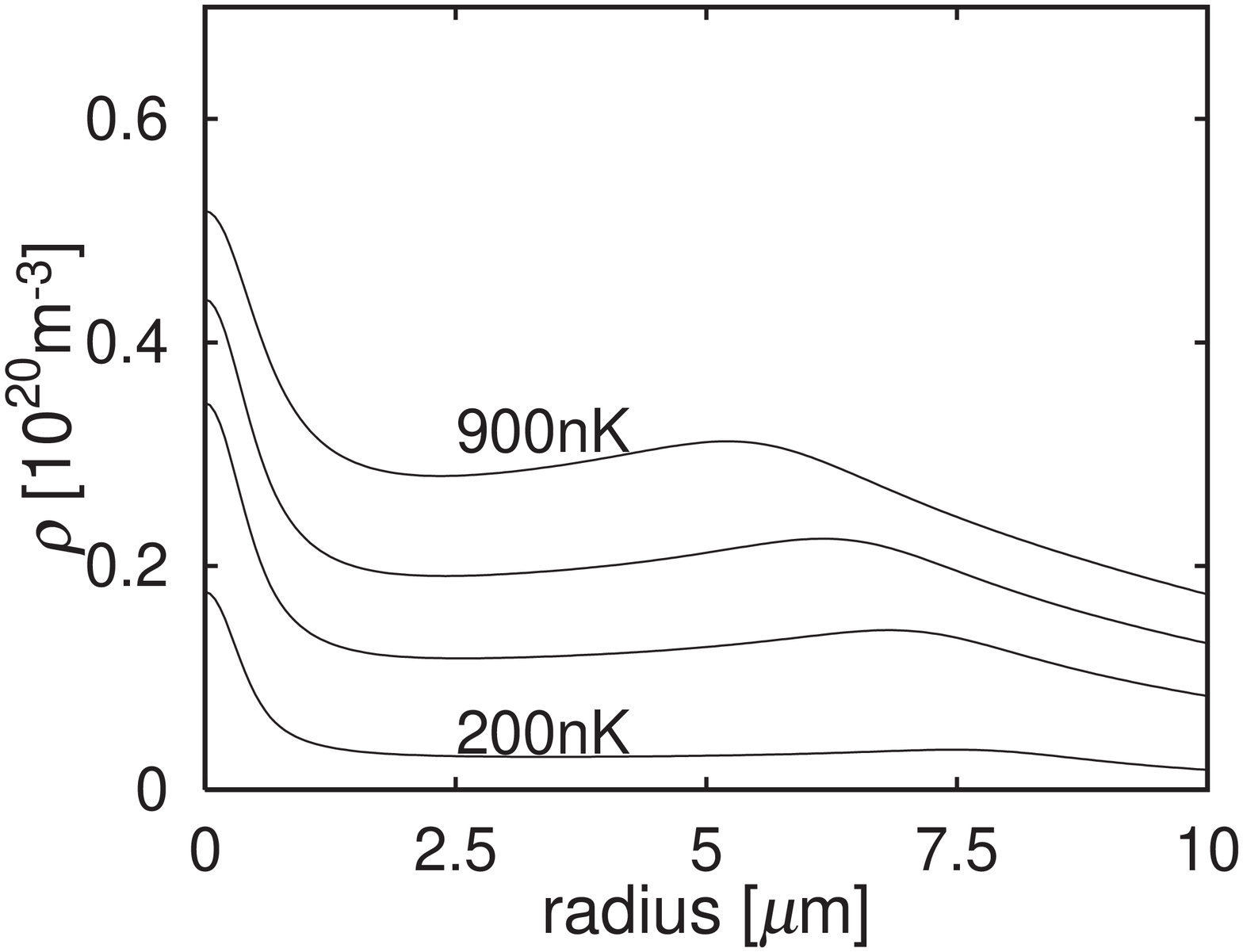}
\hspace{3.5cm}(b)

\epsfxsize=7cm
\epsfbox{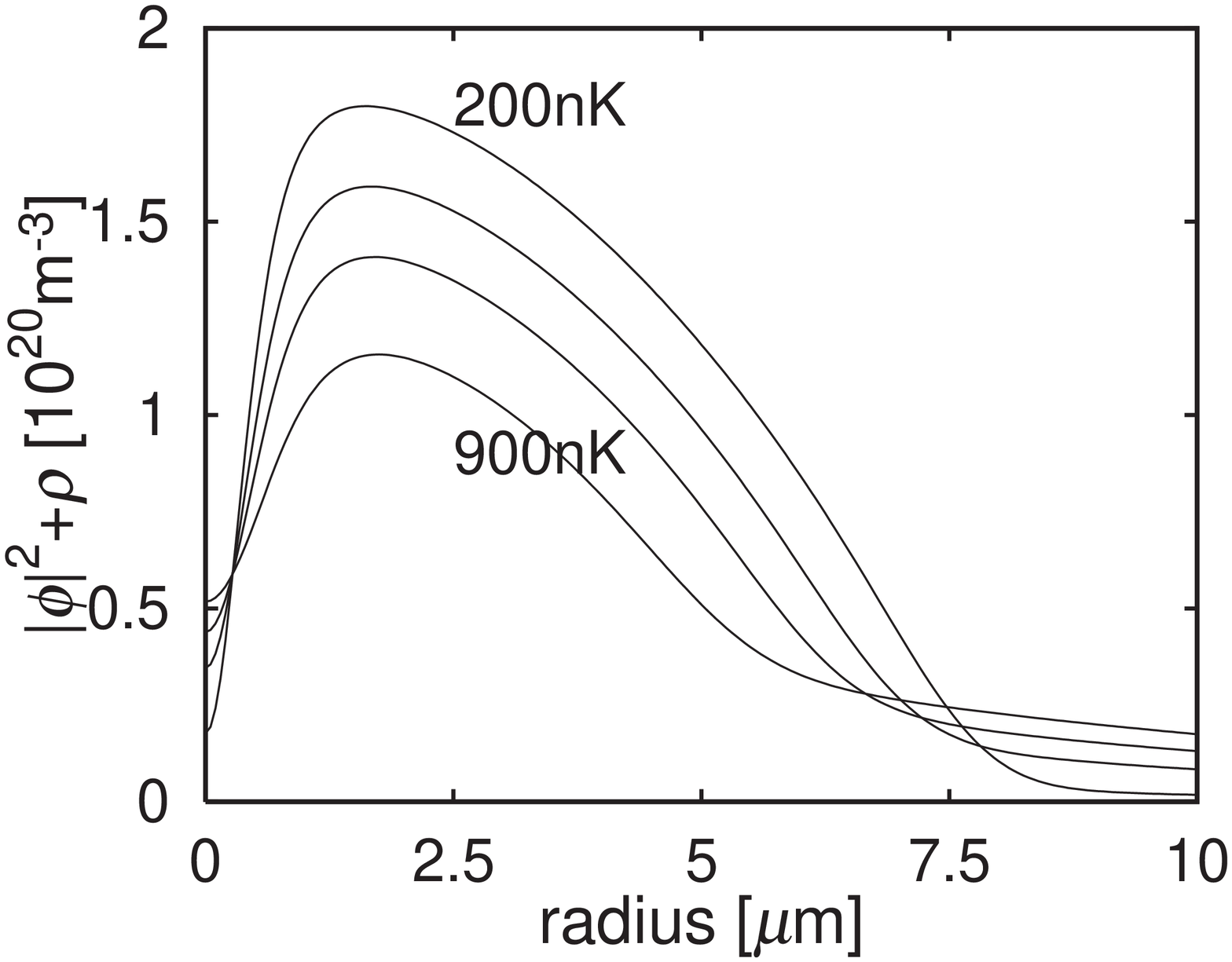}
\hspace{3.5cm}(c)

\caption{ 
(a) The condensate density $|\phi(r)|^2$ at various $T$.
The total number of the condensate $|\phi(r)|^2$ decreases as $T$ increases.
Inset shows the profile of $|\phi(r)|^2$ at small $r$.
(b) The non-condensate density $\rho(r)$.
It grows as $T$ increases.
The characteristic length scale near the center
corresponds to the core radius of $|\phi|^2$ in (a).
(c) The total density $|\phi(r)|^2 + \rho(r)$.
Each figure shows the densities at $T=200, 500, 700,$ and $900\text{nK}$.
}
\label{phi}
\end{figure}

The spatial profiles of the condensate and non-condensate density for
several temperatures are shown in Figs.\ \ref{phi}(a) and (b) respectively.
It is seen from Fig.\ \ref{phi}(a) that
the condensate $|\phi(r)|^2$ vanishes and
changes quadratically ($\propto r^2$) near the vortex center $r=0$,
which is contrasted with the linear behavior ($\propto r$)
in the Gross-Pitaevskii theory Eq.\ (\ref{eq:gpo}).
This unique shape of $|\phi(r)|^2$ reflects the
rounded shape of the $2g\rho({\bf r})$ term in Eq.\ (\ref{eq:gp}).
[This $\rho({\bf r})$ is depicted in Fig.\ \ref{phi}(b).]

This feature differs also from the superconducting
case where the pair potential changes linearly\ \cite{hohen}.
As $T$ increases, the core radius increases as seen from the inset.
The condensate is pushed outwards and converted into the non-condensate.
Thus the total number of the condensate decreases
monotonically as $T$ increases.

In Fig.\ \ref{phi}(b) the spatial variation
of the non-condensate density $\rho(r)$ is displayed.
Reflecting the core structure, the non-condensate fraction accumulates
at the vortex center.
The core radius grows as $T$ increases as inspected from
Fig.\ \ref{phi}(b), which is analyzed in more detail shortly. 

The overall feature of the spatial change of the total density
consisting of the condensate and non-condensate:
$|\phi(r)|^2+\rho(r)$ is shown in Fig.\ \ref{phi}(c).
As $T$ increases, the spatial profile tends to uniformly distribute,
and the maximum position of the density from the core also increases,
indicating that the effective core radius expands.
The total density does not vanish anywhere inside the confined potential
even at the vortex core.
It is also interesting to notice the existence of a focal point
near the center which dose not move as $T$ changes.

\subsection{Excitation Spectra}

\begin{figure}
\epsfxsize=7cm 
\epsfbox{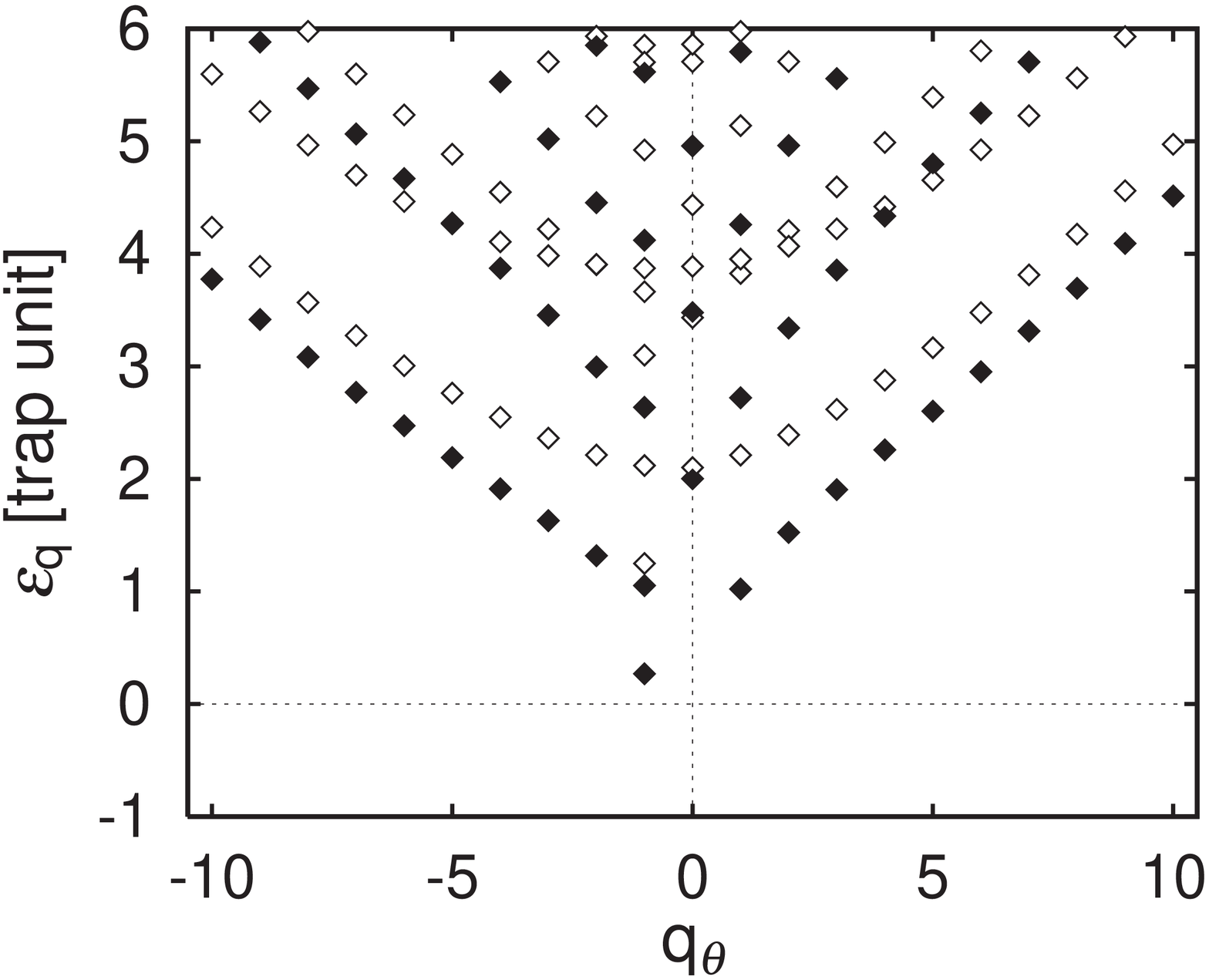}
\hspace{3.5cm}(a)

\epsfxsize=7cm 
\epsfbox{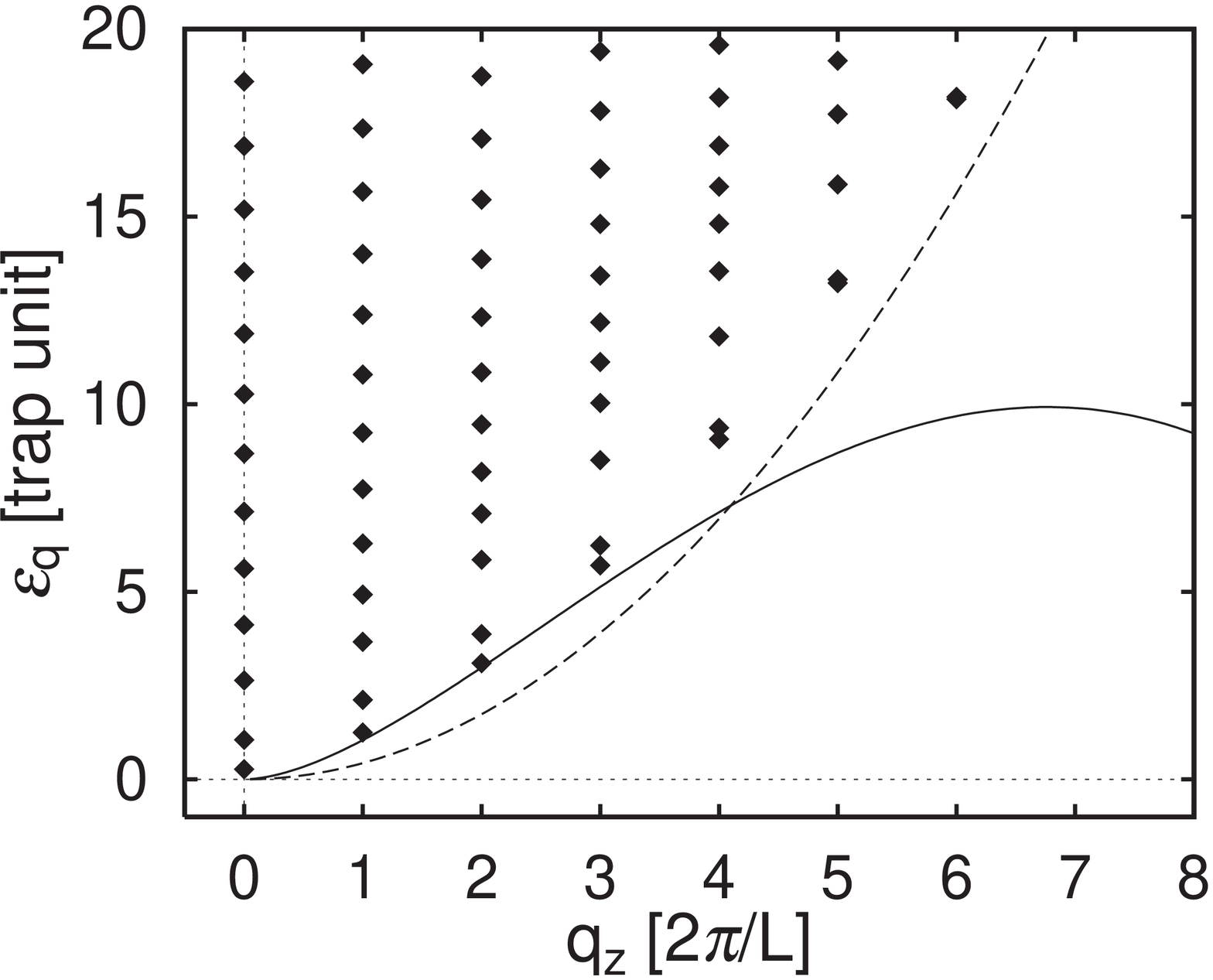}
\hspace{3.5cm}(b)

\caption{
The dispersion relations of the eigenvalues $\varepsilon_{\bf q}$
at $T=200\text{nK}$.
(a) Along $q_{\theta}$. The black points are states with $q_z = 0$.
The lowest
black
point at each $q_{\theta}$ corresponds to $q_r=1$, and
the second lowest to $q_r=2$, and so on.
(b) The eigenvalues along $q_z$. 
Only the $q_{\theta} = -1$ states are plotted.
The solid curve is explained in the text
below the Eq.\ (\protect\ref{eq:xiuni}).
The dashed curve denotes \protect$\frac{\hbar^2}{2m}q_z^2$.
}
\label{eigenvalue}
\end{figure}

In Figs.\ \ref{eigenvalue}(a) and \ \ref{eigenvalue}(b)
we display the dispersion relations
of the eigenvalues along $q_{\theta}$ and $q_z$ respectively.
The excitation modes are not symmetric about
the $q_{\theta}$-axis because of the presence of the circulation.
However, at the higher energies the distribution of these eigenmodes
become symmetric.
The modes characterized by $q_{z}=0$ are shown as black dots
in Fig.\ \ref{eigenvalue}(a).
The lowest eigenmode at $q_{\theta}=-1$ is very important in
understanding the mechanism of the vortex stabilization.

The modes with $(q_r,q_\theta) = (1,-1)$ and various $q_z$
including the LCLS correspond to
the so-called Kelvin mode known in classical vortex.
In Fig.~\ref{eigenvalue}(b) we display
the dispersion relations of the eigenvalues with $q_\theta = -1$ along $q_z$.
This dispersion relation is predicted by Pitaevskii\cite{pitaevskii} 
in a spatially uniform system with density $n_{\text{uni}}$
except near the core within the BA, at the long wave length limit as
\begin{equation}
  \varepsilon (q_r=1, q_{\theta}=-1, q_z) = 
    {\hbar^2 q_z^2 \over 2m}\ln{1\over q_z\xi_{\text{uni}}}
    \  \  (q_z\xi_{\text{uni}}\ll 1),      \label{eq:kelv}
\end{equation}
which is named the Kelvin mode, where
\begin{equation}
\\
  \xi_{\text{uni}} \equiv
    \frac{\hbar}{\sqrt{2mn_{\text{uni}}g}}.       \label{eq:xiuni}
\end{equation}
The dispersion relation of the corresponding modes
with $q_{\theta}=-1$ is shown
in Fig.~\ref{eigenvalue}(b) as the lowest edge.
We have tried to fit this prediction Eq.\ (\ref{eq:kelv}) with our result
by adjusting $n_{\text{uni}}$ in $\xi_{\text{uni}}$
of Eq.\ (\ref{eq:xiuni}).
The dotted curve in Fig.~\ref{eigenvalue}(b) 
is when $n_{\text{uni}} = 0.1 \times n_{\text{peak}}$.
The fitting is rather satisfactory.

The spatial profiles of the eigenfunctions with $q_{\theta}=-1$
are depicted in Fig.\ \ref{u}.
There is no rapid spatial oscillation in these eigenfunctions
corresponding to the Friedel oscillation seen
in the Fermi systems\cite{gygi,hayashi1}.
It is also noticed that a slow oscillation of the condensate
near the vortex core seen in several Monte Carlo computations\cite{slow}
for superfluid $^4$He is absent here.
The present weakly interacting case differs from that 
in strongly interacting system in several respects.

\subsection{Vortex Stabilization at Finite $T$}

We discuss the origin of vortex stabilization.
Let us start with the particle number density
$|\phi(r)|^2$ (condensate) and $\rho(r)$ (non-condensate)
in Fig.\ \ref{phi}.
It is the eigenstates that compose the whole system.
The wavefunction $\phi(r)$ with the energy $\mu$
corresponds to the condensate fraction and
the non-condensate density $\rho(r)$ is composed of
the eigenstates with the eigenvalues and the wavefunctions
$\varepsilon_{\bf q}$, $u_{\bf q}(r)$, and $v_{\bf q}(r)$.
The eigenstate with $\varepsilon_{\bf q}=0$ which equals $\mu$
corresponds to the condensate.

\begin{figure}
\epsfxsize=7cm
\epsfbox{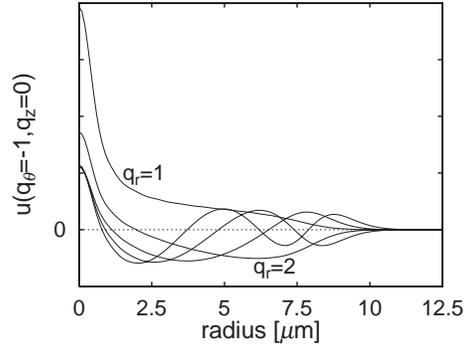}
\caption{The wave functions $u(r)(q_{\theta}=-1,q_z=0)$ at $T=200 \text{nK}$.
The radial quantum number $q_r$ varies from 1 to 5.
Each $u(r)$ has node(s) whose number is equal to $q_r$.
These states are localized at the core.
The vertical axis is arbitrary unit.
}
\label{u}
\end{figure}

The states with the angular momentum index $q_{\theta}=-1$,
whose relative motion
to the circulating current of the condensate
is at rest in the laboratory frame
[$q_{\theta} + w$ in Eq.\ (\ref{eq:phase:u}) is equal to zero],
are likely to have the lowest eigenvalue among various states.
It is the eigenstate ${\bf q} \equiv (q_r,q_{\theta},q_z) = (1,-1,0)$
whose eigenvalue becomes negative and leads to the vortex instability.
The spatial profiles of the eigenfunctions ${\bf q}=(1 \text{ to } 5,-1,0)$
have depicted in Fig.~\ref{u}.
We can see from  it that
these states are localized near the core ($r=0$).
Black dots at $q_\theta = -1$ in Fig.~\ref{eigenvalue}(a)
correspond to the wavefunctions in Fig.~\ref{u}.
We can see that the lowest mode ${\bf q}=(-1,0,1)$
is strongly localized at $r=0$.
Let us call this state the lowest core localized state (LCLS).
The vortex state is (meta)stable as far as
the LCLS exists with a positive eigenvalue.
Now the question is why the energy of the LCLS is
positive at finite $T$ as shown in Fig.~\ref{eigenvalue}(a).

We define the effective potential
\begin{equation}
  V_{\text{eff}}(r) \equiv V(r)+2g(|\phi(r)|^2 + \rho(r)), \label{eq:veff}
\end{equation}
whose combination appears
in the eigenvalue equations Eqs.\ (\ref{eq:bg1}) and (\ref{eq:bg2}).

\begin{figure}
\epsfxsize=7cm
\epsfbox{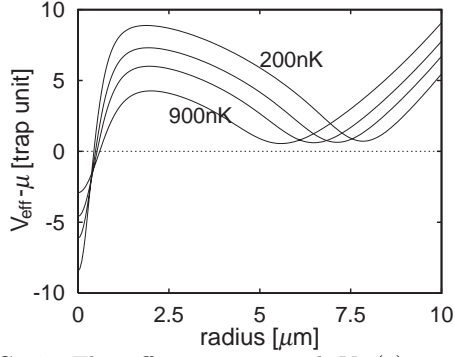}
\caption{The effective potential $V_{\text{eff}}(r)$ at
temperatures $T=200, 500, 700,$ and $900 \text{nK}$.
We subtract chemical potential $\mu$
from $V_{\text{eff}}(r)$ of each $T$.}
\label{veff}
\end{figure}

The eigenvalues of various modes are estimated
by integrating $V_{\text{eff}}(r)$ with each of the wavefunction
over the spatial volume.
Of course the LCLS is not an exception.
The shape of $V_{\text{eff}}(r)$ near the core $(r \sim 0)$ determines
the eigenvalues of the LCLS. 
As seen from Fig.~\ref{veff}, $V_{\text{eff}}(r)$ at the vortex center
is pushed upward as $T$ increases.
This is why the LCLS has a positive eigenvalue.
Although the vortex stability seems to increases with $T$,
to determine the stability conditions is a complex problem
and is discussed in the next subsection.
The increase of $V_{\text{eff}}(r)$ near the core is
caused by the contribution from the non-condensate $\rho(r)$.
Thus $V_{\text{eff}}(r)$, especially the term $2g\rho(r)$
in Eq.\ (\ref{eq:veff}),
works as an effective pinning potential
which prevents the vortex from moving outwards.
Then let us consider why the non-condensate $\rho(r)$ has a peak at the core.

\begin{figure}
\epsfxsize=8cm
\epsfbox{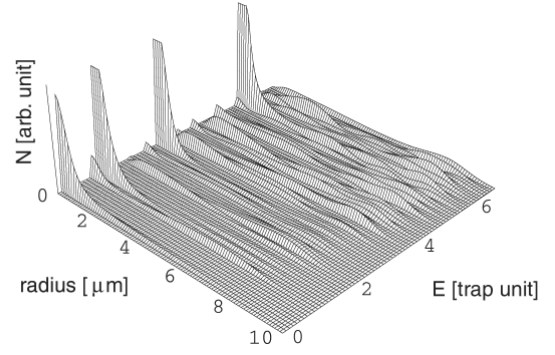}
\caption{
The local density of states $N(r,E)$
with the lower energy side ($0<E<7$ in trap unit) at $T=200\text{nK}$.
The states distinctively localized near the core
correspond to the angular momentum $q_{\theta}=-1$.
The sharp peak at the left (low energy) side is the LCLS.
Some too high peaks around the core are cut in a certain value of $N(r,E)$.
}
\label{dos}
\end{figure}

The non-condensate density $\rho(r)$ is composed of the wavefunctions 
$u(r)$ and $v(r)$ as seen from Eq.\ (\ref{eq:rho}).
Note that only $u(r)(q_{\theta}=-1)$ and
$v(r)(q_{\theta}=1)$ are non-vanishing at $r=0$ as mentioned before.
The localization of $\rho(r)$ at small $r$ mainly comes from
the LCLS term [${\bf q} = (1, -1, 0)$ term] in Eq.\ (\ref{eq:rho}).
Fig.~\ref{u} again shows that $|u(r)|^2$ coefficient of this term is
strongly localized at $r=0$, and Fig.~\ref{eigenvalue}(a) shows that
$f(\varepsilon)$ coefficient has huge value 
because the eigenvalue $\varepsilon$ is very small.
Note that $f(\varepsilon)$ is the Bose distribution function and
$\lim_{\varepsilon \to 0}f(\varepsilon) = +\infty$.
The LCLS, whose eigenvalue is approaching zero from above,
itself forms a pinning potential to keep the eigenvalue positive
combined with the Bose distribution function.
We can see the wavefunctions and the eigenvalues in Fig.~\ref{dos},
which displays the local density of states defined by Eq.\ (\ref{eq:dos}).
This may be helpful to understand the 
dominance of the LCLS in $\rho(r)$ at small $r$.

This mechanism of the $T$-dependent stabilization
is based on the PA,
which introduces the non-condensate density $\rho(r)$ as an effective potential
ignoring the friction between the condensate$(w=1)$ and
the pinning part of the non-condensate [$(q_\theta + w) = 0$].
In other words, the mechanism is based on superfluidity.

It is also obvious that
large enough $V_{\text{eff}}(r)$ near the center
is decisive in stabilizing the vortex,
irrespective of the origin either coming from
$g\rho(r)$ or from $V(r)$.
The deformation of $V(r)$ also stabilizes the vortex state and
depend neither on the PA nor on the friction between $\rho(r)$ and $\phi(r)$.
This deformation of $V(r)$ is discussed in the next section, and
the rest of this section is devoted to the various properties
of the $T$-stabilized system which has been obtained.

\subsection{Stability}

\begin{figure}
\epsfxsize=7cm
\epsfbox{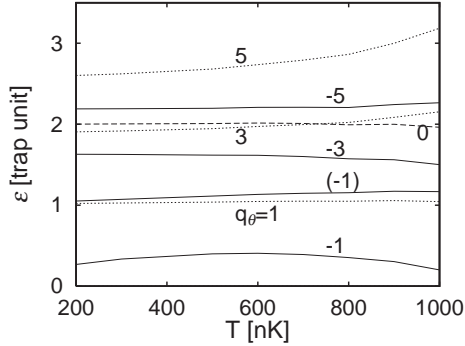}
\caption{
The eigenvalues $\varepsilon$ of the selected lowest modes
as a function of $T$.
The number of each line denotes $q_{\theta}$.
All of the lines are eigenvalues with $(q_r = 1, q_z = 0)$,
except the line labeled ``$(-1)$" with ${\bf q}=(2,-1,0)$.
Solid lines mean $q_{\theta} < 0$, dotted lines $q_{\theta} >0$, and the
dashed line $q_{\theta} = 0$.
While the dipole mode with $q_{\theta}=1$ is
exceptionally almost $T$-independent and fixed to $\varepsilon=1$,
coinciding with the trapped frequency,
other modes with $q_{\theta}>0$ ($q_{\theta}\le 0$) tend to
increase (decrease) with increasing $T$.
The eigenvalues with large $q_{\theta}$ which are not shown here
are generally increase.
This is because these wavefunctions generally lie at large radius, and
the effective potential \protect$V_{\text{eff}}(r)$ there increases
as increasing $T$.
See the large radius side of Fig.~\protect\ref{veff}.
}
\label{ET}
\end{figure}

As mentioned before,
to determine the stability conditions is a complex problem.
It is natural to consider the
eigenvalue $\varepsilon > 0$ for LCLS
and large enough $V_{\text{eff}}(r)$ at $r=0$ as possible conditions
for the stability.
What makes the matter complicated is a relation of these two conditions.
To express more explicitly,
\begin{eqnarray}
  V_{\text{eff}}(0) &=&
    g|u_{\text{LCLS}}(0)|^2 f(\varepsilon_{\text{LCLS}}) \nonumber
\\ &&
    + \text{ other contributions from} \nonumber
\\ &&
    \ u(q_\theta=-1)\text{'s and }v(q_\theta=1)\text{'s}.
\label{eq:irony}
\end{eqnarray}
The $T$-dependence of the selected lowest modes are displayed in Fig.~\ref{ET}.
As anticipated from the $T$-dependence of the lowest mode,
which gradually decreases on lowering $T$,
the whole calculation breaks down when this mode becomes negative
from above.
But this lowering process of $\varepsilon$ with $T$ will
not proceed smoothly because the Bose distribution function
at small $\varepsilon$ could make the pinning part of $V_{\text{eff}}(r)$
very large.

When $T=0$, the effective potential at the center
\begin{equation}
V_{\text{eff}}(0) =
  g \sum_{{\bf q}(q_{\theta}=1)} |v_{\bf q}(0)|^2
\end{equation}
is not likely to become large enough, although
we cannot exclude its possibility completely.

We only point out that these problems are open
and shall not discuss them further in this paper.
We failed to determine the self-consistent 
density profiles $\rho(r)$ and $|\phi(r)|^2$ for $T < 200 \text{nK}$,
which is the lowest temperature we can reach.

\subsection{Core Radius}

\begin{figure}
\epsfxsize=7cm
\epsfbox{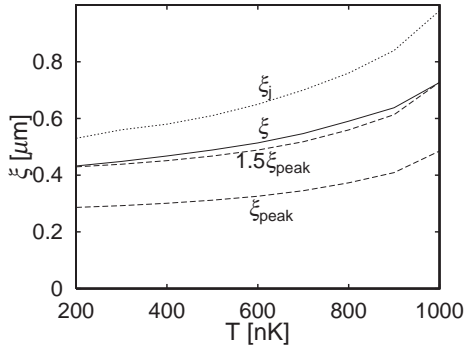}
\caption{$T$ dependencies of $\xi$'s in various definitions.
Both $\xi$ (solid line) and $\xi_{\text{peak}}$ (lower dashed line)
are defined in the text.
The upper dashed line means $1.5\times \xi_{\text{peak}}$.
The dotted line labeled $\xi_j$ is the peak radius of $j_{\theta}(r)$.
}
\label{xi}
\end{figure}

Figures~\ref{phi}(a), (b), and (c) show that
the core radius increases as $T$ increases. 
We define here the core radius
by the distance $\xi$ from the center where
 ${1\over 2}\max (|\phi(r)|^2) = |\phi(\xi)|^2$.
The $T$-dependence of this $\xi$ is exhibited in Fig.~\ref{xi}.
The core radius $\xi(T)$ is seen to be 
$T$-dependent and decrease as $T$ is lowered.
We define
$\xi_{\text{peak}} \equiv \hbar/\sqrt{2mn_{\text{peak}}g}$
where $n_{\text{peak}} \equiv \max (|\phi(r)|^2)$
is the peak density of the condensate.
This resembles the classical definition\cite{fetter}
of the coherence length $\xi_{\text{uni}}$ in Eq.\ (\ref{eq:xiuni})
for a uniform system.
The core radius $\xi$ is almost equal to
$1.5 \times \xi_{\text{peak}}$ as seen from Fig.~\ref{xi}.

The total current density $j^{(1)}_{\theta}(r) + j^{(2)}_{\theta}(r)$
defined in Eq.\ (\ref{eq:j0})
is shown in inset of Fig.~\ref{j}.
The distance from the vortex center
at which the current amplitude takes a maximum
is another measure of the vortex core radius.
This quantity $\xi_j$ shown in Fig.~\ref{xi} also decreases upon lowering $T$.

\subsection{Local Density of States}

The local density of states $N(r,E)$ 
given by Eq.\ (\ref{eq:dos}) is shown in Fig.~\ref{dos}.
The states with $q_{\theta}=-1$ are seen to localize
at the vortex center, giving rise to several prominent peaks
in the local density of states.

These features around the core may be observable and should
be compared with those in vortices in a superconductor
where extensive studies have been compiled\cite{hohen}.
This is particularly true for the localized excitations near a 
vortex core studied by Caroli {\it et al.}\cite{caroli,bardeen}.
These core excitations are now recognized to be decisive
in understanding the fundamental properties of the vortex.
The Popov theory just corresponds to the so-called 
Bogoliubov-de Gennes theory in their mathematical structures,
which is widely used
to treat various spatially non-uniform superconductors 
having interface or vortex\cite{gygi,hayashi1}.

We also note that in superconductors $N(r,E)$ is directly observed
by scanning tunneling microscope\cite{hess} and
analyzed theoretically within the similar
theoretical framework~\cite{gygi,hayashi1,hayashi2} quite successfully.

The characteristic core radius is estimated as the 
coherence length $\sim 0.5 \mu\text{m}$ which is
contrasted with a few $\AA$ in superfluid $^4$He.
Thus there is a good chance to investigate
the detailed core structure of the present BEC systems.

\subsection{Circulating Current Density}

\begin{figure}
\epsfxsize=7cm
\epsfbox{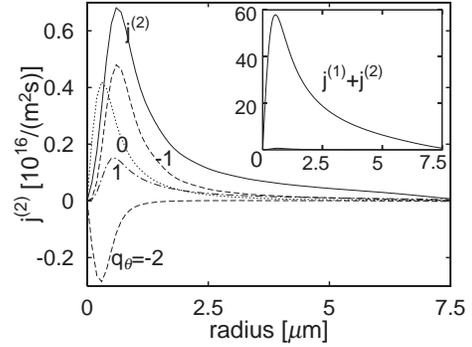}
\caption{
The non-condensate contribution $j^{(2)}_{\theta}(r)$
of the particle current density at $T=200 \text{nK}$.
Each component with $q_{\theta}=0,\pm1, -2$ is shown separately.
Inset shows the total current density
$j_{\theta}(r) \equiv j^{(1)}_{\theta}(r) + j^{(2)}_{\theta}(r)$.
}
\label{j}
\end{figure}

The total current density $j^{(1)}_{\theta}(r) + j^{(2)}_{\theta}(r)$
defined in Eq.\ (\ref{eq:j0})
is shown in inset of Fig.~\ref{j}.
The non-condensate contribution $j^{(2)}_{\theta}(r)$
and its $q_{\theta}$ components are depicted
in the main panel of Fig.~\ref{j}.
In the immediate vicinity of the core $j^{(2)}_{\theta}(r)$
is governed by the $q_{\theta}=0$ and the $q_{\theta}=-2$ components.
The positive (negative) $q_{\theta}$'s give rise to
the positive (negative) contribution to $j_{\theta}(r)$,
except for the $q_{\theta} = -1$ which has positive contribution.
The $|u(r)|^2$ term is dominant in most $q_{\theta}$ of Eq.\ (\ref{eq:j2}).
This is why positive (negative) $q_{\theta}$'s give 
the positive (negative) contribution.
Since we are treating systems with $w=1$,
the $\frac{q_{\theta} + w}{r}$ coefficients
of the $|u(r)|^2$ terms become zero
and the $|v(r)|^2$ terms with positive sign
become relevant when $q_{\theta}=-1$.
This is why the $q_{\theta} = -1$ component gives positive,
not negative, contribution.
The net result gives a finite contribution to superfluid flow
from the non-condensate.

\section{Vortex Stabilization by Pinning Potential at $T=0$}

It is now clear that introduction of a pinning potential
at the vortex center stabilizes the quantized vortex even at $T$=0.
Here we explicitly demonstrate it by a model pinning potential described by 
\begin{equation}
V_{\text{pinning}}(r)=V_0e^{-{r^2\over 2r_0^2}}  \label{vpin}
\end{equation}
in addition to the harmonic potential.
This pinning potential may be realized
by focusing a laser beam on the vortex center.
Since we are imposing the periodic boundary condition along the $z$-axis,
topologically the system is equivalent to a toroidal geometry
if the system length $L$ is large enough.

\begin{figure}
\epsfxsize=7cm 
\epsfbox{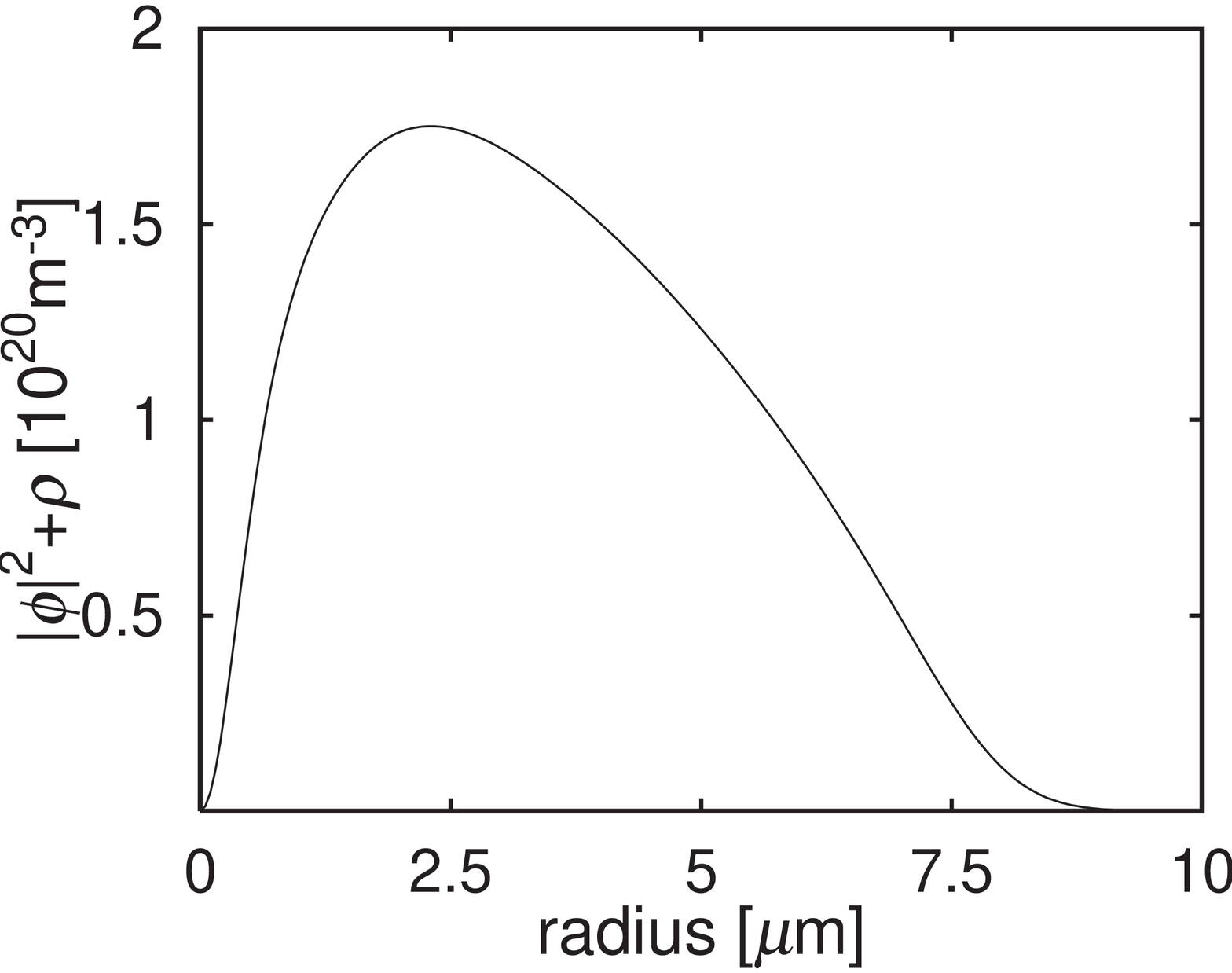}
\hspace{3.5cm}(a)

\epsfxsize=7cm
\epsfbox{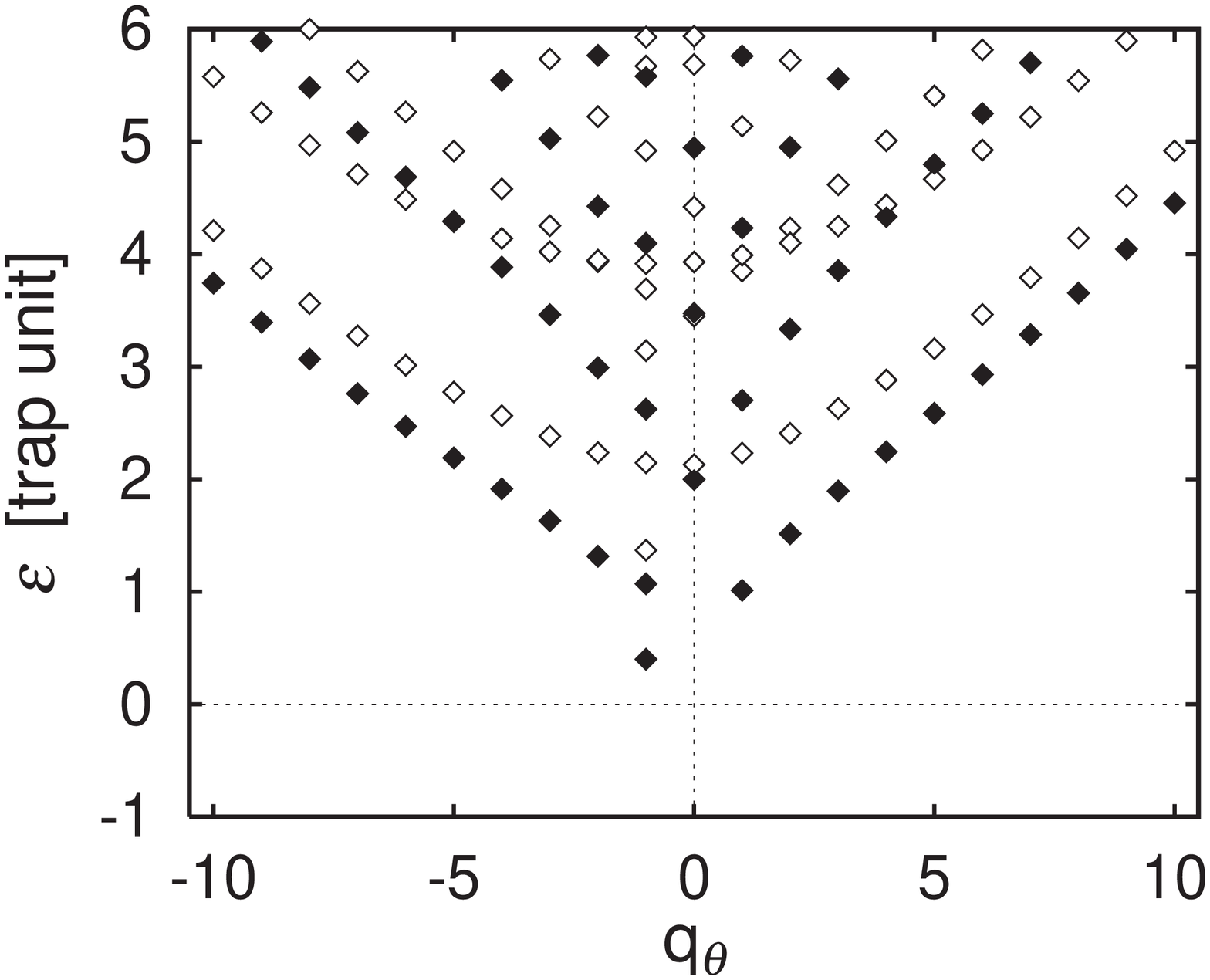}
\hspace{3.5cm}(b)

\epsfxsize=7cm
\epsfbox{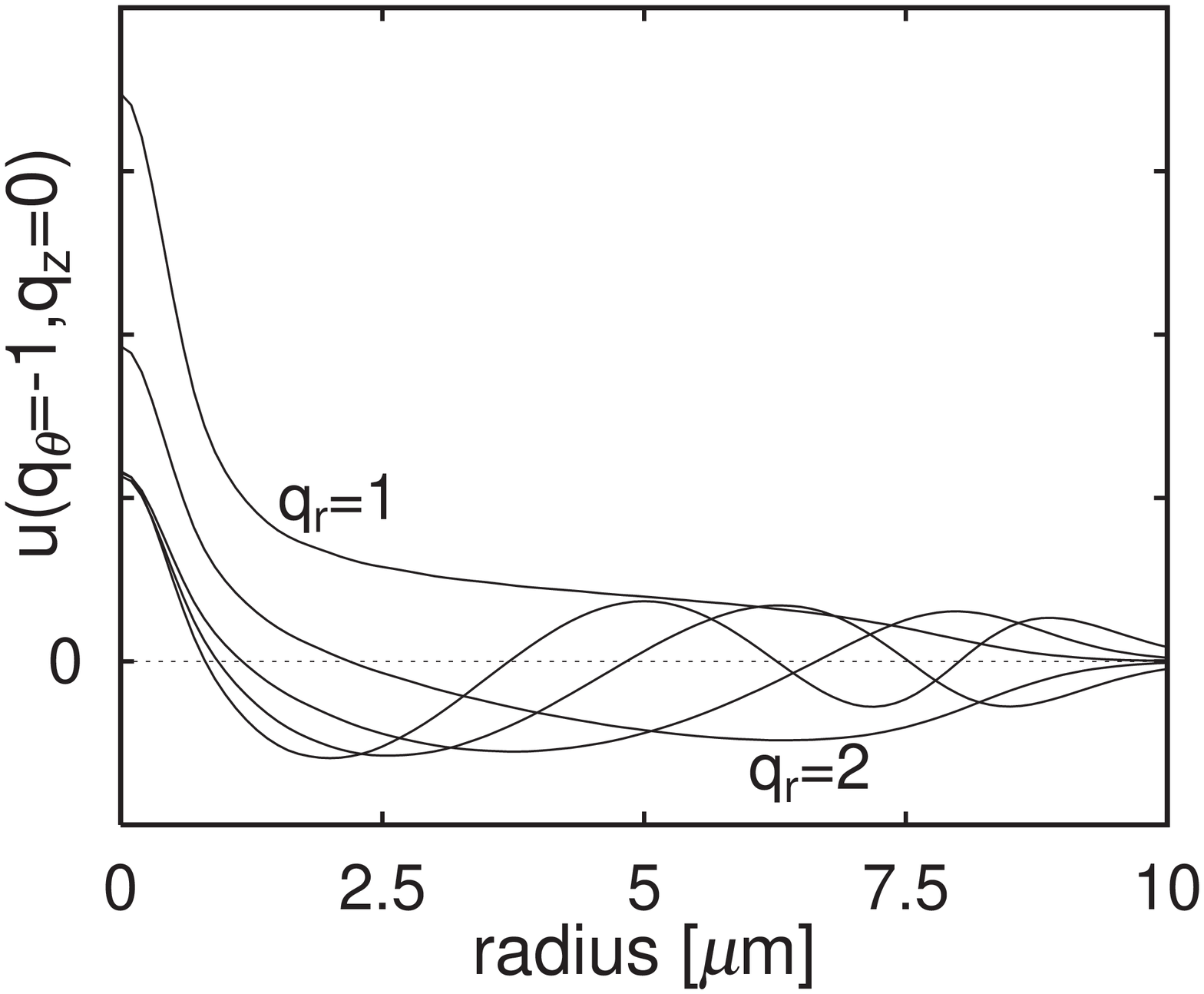}
\hspace{3.5cm}(c)

\caption{Various properties of a vortex system
stabilized by the pinning potential in Eq.\ (\protect\ref{vpin}).
$V_0=5\ \text{(trap unit)}$, $r_0=1\mu\text{m}$, and $T=0\text{K}$.
(a) The total particle number density.
(b) The dispersion relations of the eigenvalues $\varepsilon$
along $q_{\theta}$.
Black dots are eigenvalues with $q_z= 0$.
(c) The wave functions $u(r)(q_r = 1\text{ to }5,q_{\theta}=-1,q_z=0)$.
The vertical axis is arbitrary unit.}
\label{laser05}
\end{figure}

We have done self-consistent calculations at $T=0$
under the presence of the pinning potential
as the same method as in the previous section.
The self-consistent solution for $V_0=5$ trap unit and $r_0=1\mu \text{m}$
and the other parameters are same as before
is depicted in Figs.~\ref{laser05}(a), (b), and (c).
It is seen from those that the vortex is stable
because the LCLS is now all positive
as shown in Fig.~\ref{laser05}(b),
allowing the fully self-consistent solution.
We note that the eigenfunction $u(r)(q_r=1)$ shown in Fig.~\ref{laser05}(c)
has still large amplitude at $r=0$.
The condensate is pushed outwards
because of the pinning potential,
and its initial rise at the vortex core is steep [Fig.~\ref{laser05}(a)].
The non-condensate fraction is vanishingly small in this example at $T=0$.

\begin{figure}
\epsfxsize=7cm 
\epsfbox{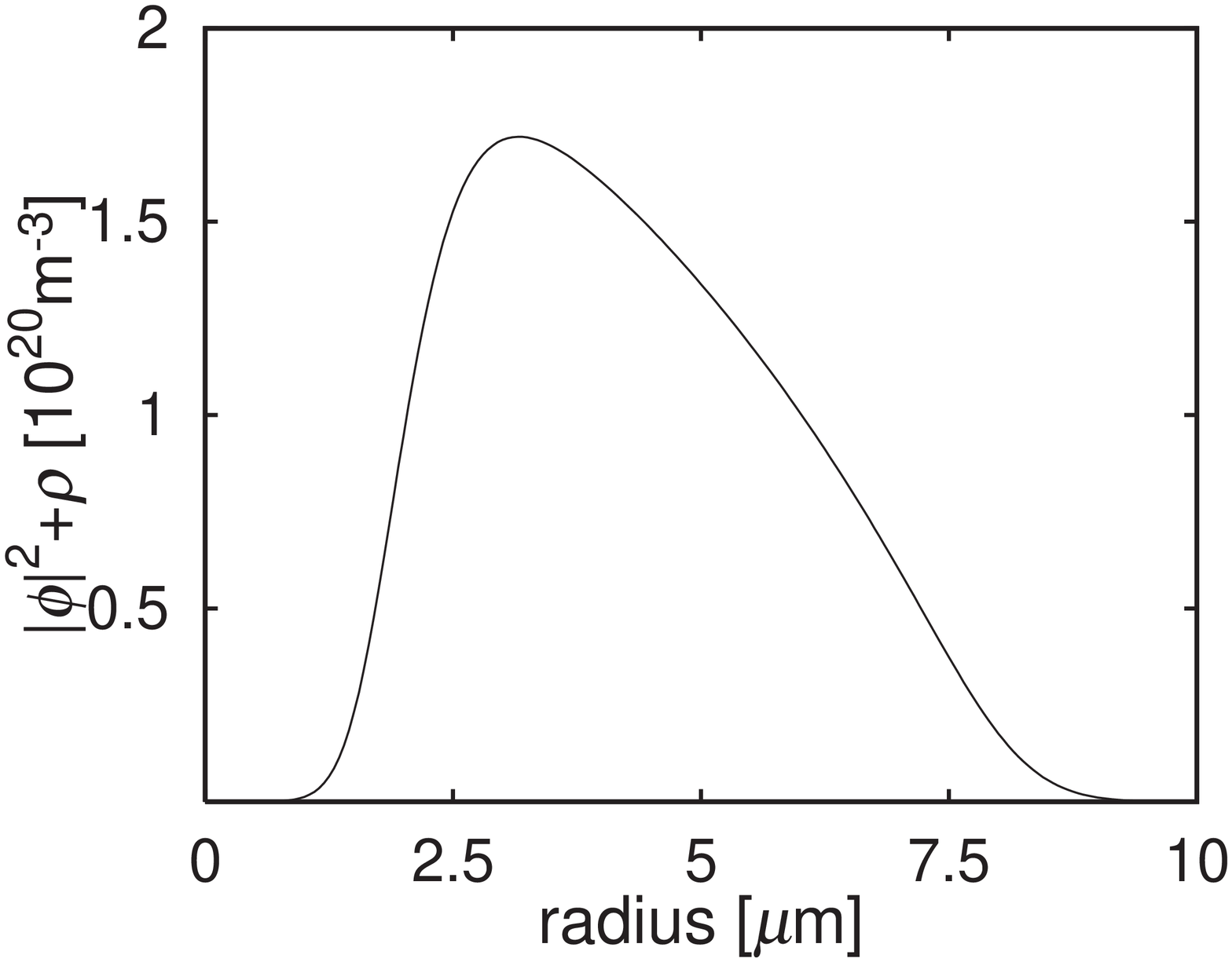}
\hspace{3.5cm}(a)

\epsfxsize=7cm 
\epsfbox{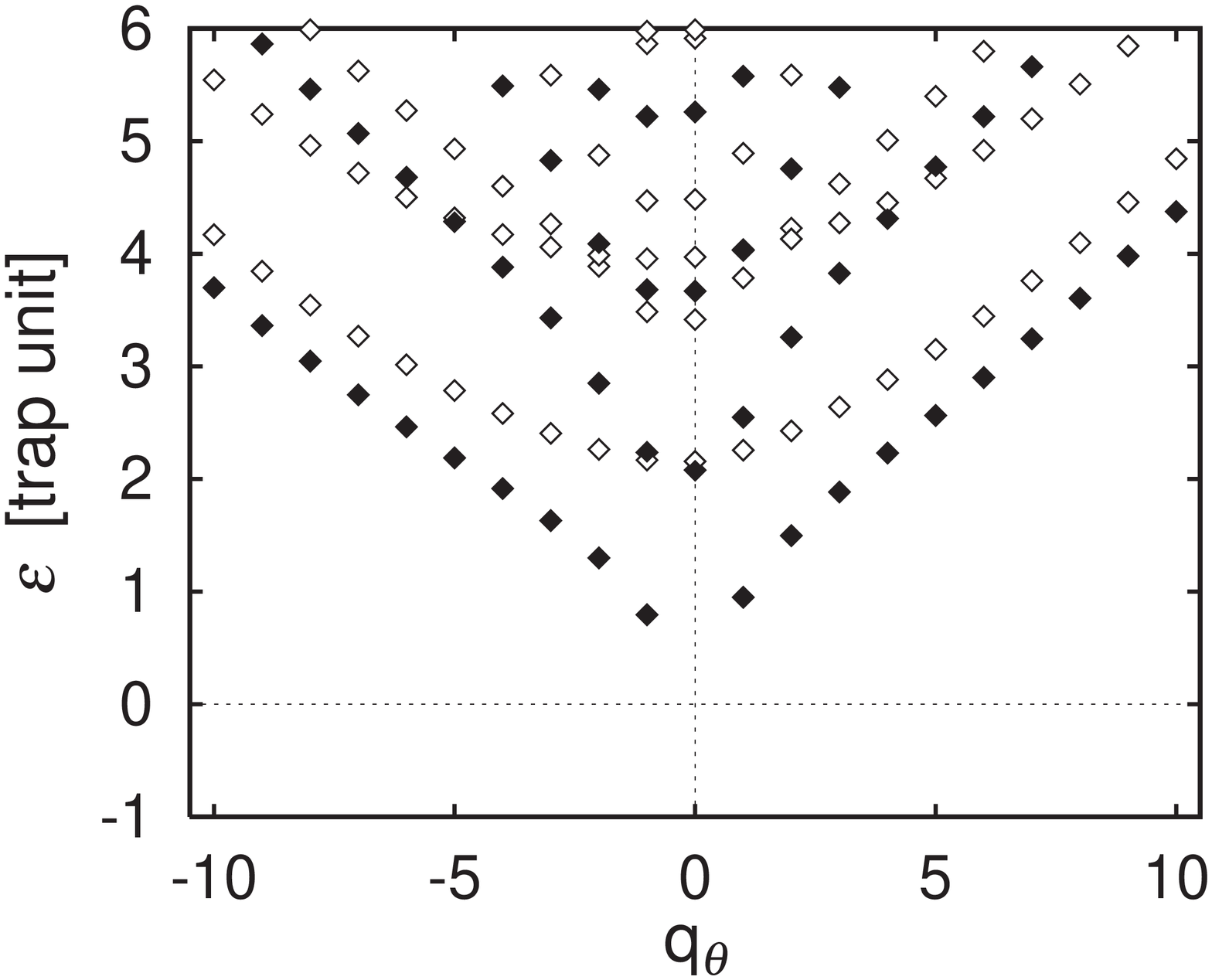}
\hspace{3.5cm}(b)

\epsfxsize=7cm 
\epsfbox{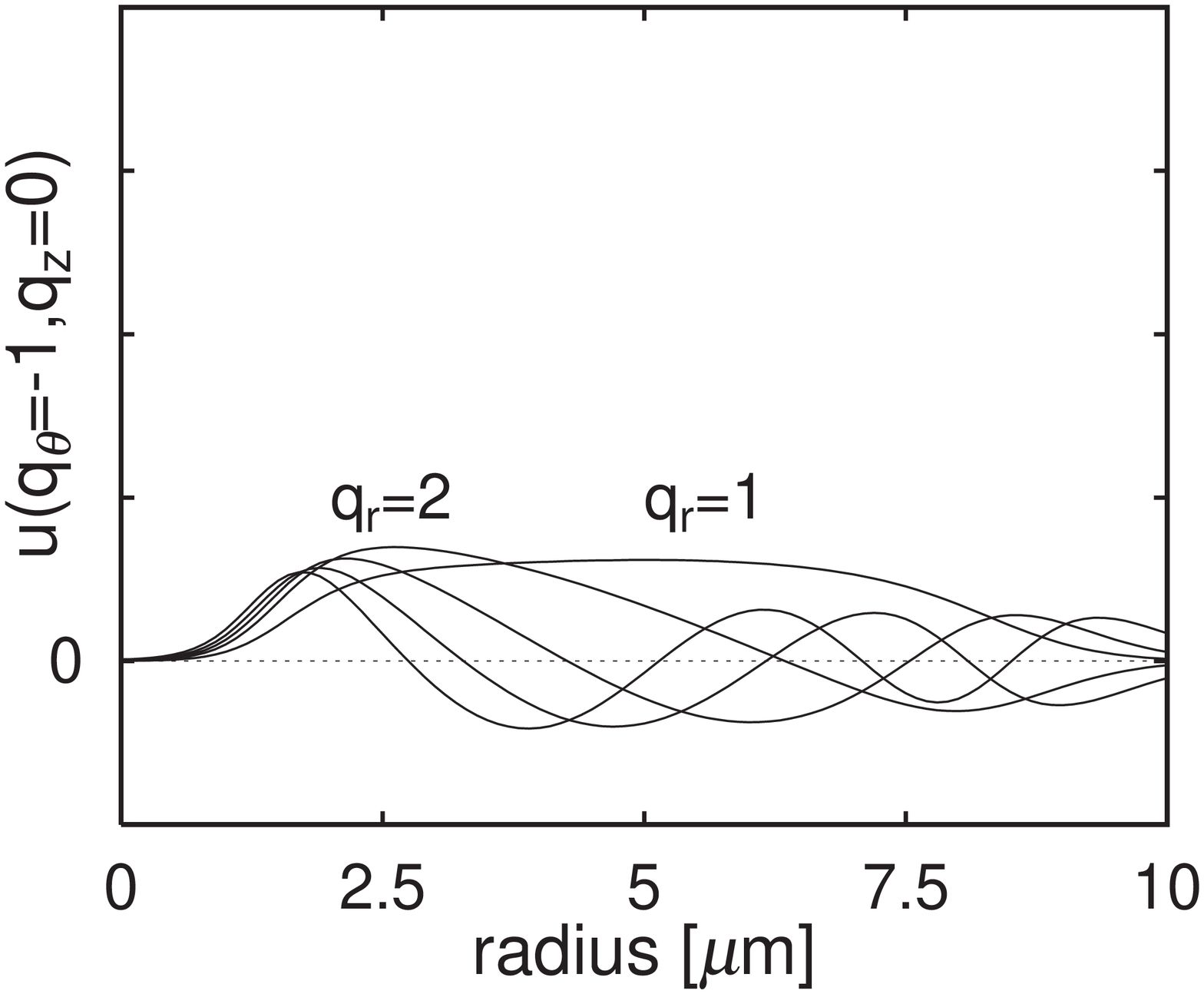}
\hspace{3.5cm}(c)

\caption{
Various properties of a vortex system stabilized by the pinning potential.
$V_0=50\ \text{(trap unit)}$, $r_0=1\mu\text{m}$, and $T=0\text{K}$. 
(a) The total particle number density.
(b) The dispersion relations of the eigenvalues $\varepsilon$
along $q_{\theta}$. Black dots are eigenvalues of $q_z= 0$.
(c) The wave functions $u(r)(q_r = 1\text{ to }5,q_{\theta}=-1,q_z=0)$.
The vertical axis is arbitrary unit.}
\label{laser50}
\end{figure}

In a larger pinning potential ($V_0=50$ trap unit and $r_0=1\mu \text{m}$)
exemplified by Fig.~\ref{laser50} the condensate is farther pushed outwards
and the vortex center is empty space [Fig.~\ref{laser50}(a)].
It resembles a ring-shaped BEC systems which has been considered
theoretically\cite{rokhsar,ring}. 
The wave functions of the states ${\bf q}=(1 \text{ to }5,-1,0)$
in Fig.~\ref{laser50}(b) is squeezed at small $r$ by the pinning potential. 
There is no LCLS.
This deformation of $u(r)$ directly lifts the eigenvalue of
the lowest eigenstate at $q_{\theta}=-1$ in Fig.~\ref{laser50}(b) 
to stabilize further the vortex state.
The overall distribution of the excitation spectrum is
almost symmetric about the $q_{\theta}$ axis [Fig.\ \ref{laser50}(b)].
All the corresponding eigenfunctions
with $q_{\theta}=-1$ vanish at $r=0$ [Fig.~\ref{laser50}(c)],
which is contrasted with those in Fig.~\ref{u} and Fig.~\ref{laser05}(c).

\section{Conclusion and Discussions}

We have investigated the stability conditions of a vortex in BEC systems
by performing microscopic self-consistent calculations
based on the Popov approximation.
It is demonstrated that the BEC systems confined
in a harmonic potential at finite temperatures sustain the stable vortex.
This is contrasted with the cases
in the absence of
the pinning potential
where the same calculation
yields unstable vortex as shown previously\cite{isoshima,rokhsar}.
The vortex instability comes from
the negative eigenvalue of the LCLS (lowest core localized state).
Therefore physically it is possible to lift this energy
by increasing the effective potential felt by this quasi-particle.
This can be realized by raising $T$ or introducing
an external pinning potential.
In the former case the 
LCLS itself acts as the effective pinning potential,
amplified through the Bose distribution function.
The nature of the frictionless interaction
between the condensate and the non-condensate in the Popov theory
enables this $T$-dependent mechanism.

The detailed vortex properties are analyzed for the stable vortex.
Several eminent features are worth mentioning:
(1) The condensate $|\phi (r)|^2$ behaves as $\propto r^2$
near the vortex core and the non-condensate has a peak at the core.
(2) The characteristic length or the core radius
is well described by a formula $\hbar/\sqrt{2mn_{\text{peak}}g}$
with $n_{\text{peak}}$ being the peak density of the condensate. 
(3) The particle density is non-vanishing everywhere
inside a system even at the core center when a system is $T$-stabilized.
(4) The local density of states exhibits large peak structures
at the core center for the low energy side,
and the peaks mean the existence of the LCLS and other core localized
states. These characteristics may be directly observed
once the quantized vortex is realized in the present BEC systems.

We hope that the vortex state in BEC of alkali atom gas
should come true experimentally.
If realized, we may have a chance to unify theories of the
present dilute BEC systems and the strongly interacting
superfluid $^4$He systems.
We also hope that the precise and detailed analysis of
the vortex core structure should be possible
because an advantage here is that the length scale of the core radius is
much larger than those in superfluid $^4$He or superconductors.

\section*{acknowledgments}
The authors thank T. Ohmi, D. S. Rokhsar, A. L. Fetter,
Y. Takahashi, and K. Toyoda
for useful discussions.
In particular, the communication with D. S. Rokhsar certainly encourages us 
to perform this study.
Some numerical computation in this work was carried out
at the Yukawa Institute Computer Facility.


\newpage

\newpage

\newpage

\newpage

\begin{references}

\bibitem[*]{tomoya}
Electronic address: tomoya@mp.okayama-u.ac.jp

\bibitem{cornell}M. H. Anderson, J. R. Ensher, M. R. Matthews,
C. E. Wieman, and E. A. Cornell, Science {\bf 269}, 198 (1995).

\bibitem{hulet}C. C. Bradley, C. A. Sackett, J. J. Tollett, and R. G. Hulet,
Phys. Rev. Lett. {\bf 75}, 1687 (1995).

\bibitem{ketterle}K. B. Davis, M.-O. Mewes,
M. R. Andrews, N. J. van Druten, D. D. Durfee, D. M. Kurn, and W. Ketterle,
Phys. Rev. Lett. {\bf 75}, 3969 (1995).

\bibitem{review2}{\it Bose-Einstein Condensation},
A. Griffin, D. W. Snoke, and S. Stringari, Eds.
(Cambridge University Press, Cambridge, England, 1995).

\bibitem{review1}See for review,
F. Dalfovo, S. Giorgini, L. P. Pitaevskii, and S. Stringari,
preprint(cond-mat/9806038).

\bibitem{feshbach}S. Inouye, M. R. Andrews, J. Stenger, H.-J. Miesner,
D. M. Stamper-Kurn, and W. Ketterle, Nature {\bf 392}, 151 (1998).

\bibitem{kurn}D. M. Stamper-Kurn, M. R. Andrews, A. P. Chikkatur,
S. Inouye, H.-J. Miesner, J. Stenger, and W. Ketterle,
Phys. Rev. Lett. {\bf 80}, 2027 (1998).

\bibitem{ohmi} T. Ohmi and K. Machida,
J. Phys. Soc. Jpn. {\bf 67}, 1822 (1998).

\bibitem{ho}T. L. Ho, preprint(cond-mat/9803231).

\bibitem{laserkarl} K.-P. Marzlin and W. Zhang,
Phys. Rev. A {\bf 57}, 3801 (1998); {\bf 57}, 4761 (1998).

\bibitem{gp2d} B. Jackson, J. F. McCann, and C. S. Adams,
Phys. Rev. Lett. {\bf 80}, 3903 (1998).

\bibitem{bogoliubov}N. Bogoliubov,
J. Phys. (USSR) {\bf 11}, 23 (1947).

\bibitem{gross}E. P. Gross,
Nuovo Cimento {\bf 20}, 454 (1961); 
J. Math. Phys. {\bf 4}, 195 (1963).

\bibitem{pitaevskii}L. P. Pitaevskii,
Zh. Eksp. Teor. Fiz. {\bf 40}, 646 (1961)
[English Transl. Sov. Phys. -JETP {\bf 13}, 451 (1961)].

\bibitem{iordanskii}S. V. Iordanskii,
Zh. Eksp. Teor. Fiz. {\bf 49}, 225 (1965) 
[English Transl. Sov. Phys. -JETP {\bf 22}, 160 (1966)].

\bibitem{fetter}A. L. Fetter,
Phys. Rev. {\bf 138}, A709 (1965);
{\bf 140}, A452 (1965);
Ann. Phys. {\bf 70}, 67 (1972).

\bibitem{popov}V. N. Popov,              
{\it Functional integrals and collective excitations},
(Cambridge University Press, Cambridge, England, 1987).

\bibitem{meanfield}See for various applications of mean field theories
to the present finite systems, 
V. V. Goldman, I. F. Silvera, and A. J. Leggett,
Phys. Rev. B {\bf 24}, 2870 (1981);
D. A. Huse and E. D. Siggia,
J. Low Temp.Phys. {\bf 46}, 137 (1982);
M. Edwards and K. Burnett,
Phys. Rev. A {\bf 51}, 1382 (1995);
P. A. Ruprecht, M. J. Holland, K. Burnett, and M. Edwards,
Phys. Rev. A {\bf 51}, 4704 (1995).

\bibitem{isoshima}T. Isoshima and K. Machida,
J. Phys. Soc. Jpn. {\bf 66}, 3502 (1997).

\bibitem{dodd}
R. J. Dodd, K. Burnett, M. Edwards, and  C. W. Clark,
Phys. Rev. A {\bf 56}, 587 (1997).

\bibitem{rokhsar}D. S. Rokhsar,
Phys. Rev. Lett. {\bf 79}, 2164 (1997); preprint(cond-mat/9709212).

\bibitem{isoshima2}T. Isoshima and K. Machida,
J. Phys. Soc. Jpn. {\bf 67}, 1840 (1998).

\bibitem{gpvortex}
M. Edwards, R. J. Dodd, C. W. Clark, P. A. Ruprecht, and K. Burnett, 
Phys. Rev. A {\bf 53}, R1950 (1996);
F. Dalfovo and S. Stringari, Phys. Rev. A {\bf 53}, 2477 (1996).

\bibitem{sinha} S. Sinha, 
Phys. Rev. A {\bf 55}, 4325 (1997).
This paper treats semiclassicaly an equation similar to ours.

\bibitem{svi}A. A. Svidzinsky and A. L. Fetter, preprint(cond-mat/9803181).

\bibitem{donnelly}R. J. Donnelly,
{\it Quantized vortices in Helium II}
(Cambridge University Press, Cambridge, 1991). 

\bibitem{hohen}For a review, A. L. Fetter and P. C. Hohenberg,
in {\it Superconductivity},
Ed. by R. D. Parks (Marcell Dekker, New York, 1969).

\bibitem{hutchinson}D. A. W. Hutchinson, E. Zaremba, and A. Griffin,
Phys. Rev. Lett. {\bf 78}, 1842 (1997).

\bibitem{hess}H. Hess, R. B. Robinson, and J. V. Waszczak,
Phys. Rev. Lett. {\bf 64}, 2711 (1990). 

\bibitem{gygi}F. Gygi and M. Schl\"uter,
Phys. Rev. B {\bf 43}, 7609 (1991). 

\bibitem{hayashi1}N. Hayashi, T. Isoshima, M. Ichioka, and K. Machida,
Phys. Rev. Lett. {\bf 80}, 2921 (1998).

\bibitem{slow}
G. V. Chester, R. Metz, and L. Reatto,
Phys. Rev. {\bf 175}, 275(1968);
F. Dalfovo, Phys. Rev. B {\bf 46}, 5482 (1992).

\bibitem{hayashi2}N. Hayashi, M. Ichioka, and K. Machida,
Phys. Rev. Lett. {\bf 77}, 4074 (1996);
Phys. Rev. B {\bf 56}, 9052 (1997).

\bibitem{ring}E. J. Mueller, P. M. Goldbart, and Y. L.-Geller,
Phys. Rev. A {\bf 57}, R1505 (1998).

\bibitem{caroli}C. Caroli, P. G. de Gennes, and J. Matricon,
Phys. Lett. {\bf 9}, 307 (1964);
C. Caroli and J. Matricon,
Phys. Kondens. Materie. {\bf 3}, 380 (1965). 

\bibitem{bardeen}J. Bardeen, R. K\"ummel, A. E. Jacobs, and L. Tewordt,
Phys. Rev. {\bf 187}, 556 (1969).

\end{references}
\end{document}